\documentstyle[12pt]{article}
\let\ssection=\section
\renewcommand{\section}{\setcounter{equation}{0}\ssection}

\newcommand\mathC{\mkern1mu\raise2.2pt\hbox{$\scriptscriptstyle|$}
        {\mkern-7mu\rm C}}              % The complex  numbers
\newcommand{\mathR}{{\rm I\! R}}         % The real numbers
\newcommand{\ket}[1]{|#1\rangle}

\newcommand\mapdown[1]{\Big\downarrow
                        \rlap{$\vcenter{\hbox{$\scriptstyle#1$}}$}}
\newcommand\mapright[1]{\smash{
        \mathop{\mbox{\large{$\longrightarrow$}}}\limits^{#1}}}
\newcommand\bundle[3]{\begin{array}[t]{c}
        {#1}\\ \mapdown{#2}\\ {#3}\end{array}}
\newcommand\bundlemap[2]{\begin{array}[t]{c}
\mapright{#1}\\
\phantom{\mapdown{}}\\\mapright{#2}\\\end{array}}

\begin{document}
\begin{titlepage}
\hspace{10truecm}Imperial/TP/00-01/32

\begin{center}
{\large\bf Quantising the Foliation in History Quantum Field
Theory}
\end{center}

\vspace{0.8 truecm}

\begin{center}
        C.J.~Isham\footnote{email: c.isham@ic.ac.uk}\\[10pt]
        The Blackett Laboratory\\
        Imperial College of Science, Technology \& Medicine\\
        South Kensington\\
        London SW7 2BZ\\
\end{center}

\begin{center}
        and
\end{center}

\begin{center}
        K.~Savvidou\footnote{email: k.savvidou@ic.ac.uk}\\[10pt]
        The Blackett Laboratory\\
        Imperial College of Science, Technology \& Medicine\\
        South Kensington\\
        London SW7 2BZ\\
\end{center}

\begin{center}
27 October 2001
\end{center}

\begin{abstract}
As a preliminary to discussing the quantisation of the foliation
in a history form of general relativity, we show how the
discussion in \cite{SavQFT1} of a history version of free, scalar
quantum field theory can be augmented in such a way as to include
the quantisation of the unit-length, time-like vector that
determines a Lorentzian foliation of Minkowski spacetime. We
employ a Hilbert bundle construction that is motivated by: (i)
discussing the role of the external Lorentz group in the existing
history quantum field theory \cite{SavQFT1}; and (ii) considering
a specific representation of the extended history algebra
obtained from the multi-symplectic representation of scalar field
theory.
\end{abstract}
\end{titlepage}

\section{Introduction}\label{Sec:Introduction}
The goal of the present paper is to extend the discussion in
\cite{SavQFT1} of the construction of a history version of
quantum scalar field theory in Minkowski spacetime. In
particular, we shall show how the formalism can be developed to
include the quantisation of the four-vector $n$ that determines
the spacetime foliation that plays a central role in the theory.
The motivation for such a step, and the relevant background
information, is as follows.

The `consistent-histories' approach to quantum theory was
originally introduced to provide a novel way of re-interpreting
standard quantum theory, particularly in regard to the role
played by measurement. However, because of the novel way in which
time is handled, consistent-history theory also has the potential
for providing new and powerful ways of studying quantum theories
of gravity. Most recently, in \cite{SavGR1} the formalism was
applied to construct a history version of the canonical form of
classical general relativity. The possibility also arises to use
this formalism in the context of generalised ideas of time and
space: for example, in models where spacetime is not represented
by a differentiable manifold.

A first step in developing the framework with this goal in mind
was taken in \cite{Ish94} where a new mathematical
formalism---the `History Projection Operator' (HPO) method---was
introduced. This places emphasis on the idea of `quantum temporal
logic', and potentially allows substantial generalisations of the
notion of time. The heart of this formalism is the idea that
propositions about the {\em temporal history\/} of a system
should be represented by projection operators on a `history'
Hilbert space\footnote{This is to be contrasted with the
situation in standard quantum theory in which projection
operators represent propositions about the system at a {\em
single\/} time.}. In the case of simple, Newtonian time, and
histories labelled by a finite set of discrete time points, the
history Hilbert space is a tensor product of a copy of the
standard canonical Hilbert space for each such time point.

The idea of representing history propositions by projection
operators lead in turn to the notion of a `{\em history
group\/}'. This is the history analogue of the Weyl group and its
associated canonical commutation relations; in particular, the
spectral projectors of the history operators in the Lie algebra
of the history group represent propositions about the associated
history quantities.

The introduction of a history group is particularly useful in the
context of histories with a {\em continuous\/} time label, since
it is by no means a trivial matter to define the continuous
analogue of a tensor product. Instead, one finds the history
Hilbert space by looking for representations of the appropriate
history algebra.

For example, for the case of a point particle moving in one
dimension, the history algebra for histories labelled with a
continuous time parameter $t$ is \cite{IL95} \cite{ILSS98}
\begin{eqnarray}
    {[}\,\hat x_t,\hat x_{t'}\,]&=&0        \label{HGPPxx}\\
    {[}\,\hat p_t,\hat p_{t'}\,]&=&0        \label{HGPPpp}\\
    {[}\,\hat x_t,\hat p_{t'}\,]&=&i\hbar\tau\delta(t-t'),
                                            \label{HGPPxp}
\end{eqnarray}
and the basic history propositions in the theory refer to the
value of time-averaged quantities such as ${1\over\tau}\int dt\,
x_t\,f(t)$ and ${1\over\tau}\int dt\,p_t\,h(t)$ where $f$ and $h$
are smearing functions. Note that, in Eq.\ (\ref{HGPPxp}), $\tau$
is a new constant in the theory with the dimension of
time.\footnote{In discussions involving the use of a history
algebra with continuous time there is a tendency to choose units
in which $\tau=1$. However, this constant remains lurking in the
background.}

In equations (\ref{HGPPxx})--(\ref{HGPPxp}), the label $t$ on the
operators $\hat x_t$ and $\hat p_t$ refers to the time at which
propositions about the system are asserted---the time of
`temporal logic'. It was to include in an explicit way such a
time of temporal logic that the HPO formalism was originally
developed. However, a clear notion of dynamics was not implemented
for the, naturally time-averaged, physical quantities of the
theory.

A major advance in the HPO formalism took place when time was
introduced in a completely new way \cite{Sav99a} \cite {Sav99b}.
It was realised that it is natural to consider time in a two-fold
manner: the `time of being'---the time at which events `happen'
(and from this perspective, the time label $t$ in Eqs.\
(\ref{HGPPxx})--(\ref{HGPPxp}) and in Eq.\ (\ref{Def:xt(s)})
below can be regarded as such), and the `time of becoming'---the
time of dynamical change, represented by a time label $s$. This
{\em second\/} time appears in the history analogue $\hat x_t(s)$
of the Heisenberg picture, which is defined as
\begin{equation}
  \hat x_t(s):=e^{is\hat H/\hbar}\hat x_t e^{-is\hat H/\hbar}
                    \label{Def:xt(s)}
\end{equation}
where $\hat H:={1\over\tau}\int dt\hat H_t$ is the history
quantity that represents the time average of the energy of the
system. The notion of time evolution is now recovered for the
time-averaged physical quantities, for example
\begin{equation}
  \hat x_f(s):=e^{is\hat H/\hbar}\hat x_f e^{-is\hat H/\hbar}
                    \label{Def:xf(s)}
\end{equation}
where $f(t)$ is a smearing function.

Associated with these two manifestations of the concept of time
are two types of time transformation: the `external' translation
\begin{equation}
    \hat x_t(s)\mapsto\hat x_{t+t'}(s),      \label{ExtTTransl}
\end{equation}
and the `internal' translation
\begin{equation}
    \hat x_t(s)\mapsto\hat x_t(s+s').      \label{IntTTransl}
\end{equation}
The external time translation is generated by the `Louiville'
operator \cite{Sav99a}
\begin{equation}
    V:=\int dt\,\hat p_t {d\hat x_t\over dt}
\end{equation}
whereas the internal time translation is generated by the
time-averaged energy operator $\hat H$.

More importantly, it was shown in \cite{Sav99a} that the generator
of time translation in the HPO theory is the `action' operator
$S$ defined as
\begin{equation}
    S:=\int dt\,\hat p_t {d\hat x_t\over dt} - H = V-H.
\end{equation}
Hence the action operator is the generator of {\em both\/} types
of time translation
\begin{equation}
    \hat x_t(s)\mapsto\hat x_{t+t'}(s+s').      \label{ExtTTranslboth}
\end{equation}
It is a very striking result that in the HPO theory the quantum
analogue of the classical action functional is an actual operator
in the formalism, and is the generator of time translations
\cite{Sav99a}.

The idea of `two times'---and the associated two types of time
translation---has recently been generalised to relativistic field
theory \cite{SavQFT1} where, in particular, it is shown that the
analogue of the two types of time translation is the existence of
two {\em Poincar\'e\/} groups. The goal of the present paper is to
develop the ideas in \cite{SavQFT1} in one particular respect:
namely, the way in which spacetime foliations enter the theory.

That the idea of a Lorentzian\footnote{By `Lorentzian' we mean
that each leaf of the foliation is a hyperplane in the Minkowski
spacetime.} spacetime foliation should play an important role in
a history quantum field theory is understandable. Indeed, the
obvious analogue of the history algebra Eqs.\
(\ref{HGPPxx})--(\ref{HGPPxp}) for a quantum scalar field theory
is (choosing units from now on such that $\tau=1$)
\begin{eqnarray}
 {[\,} {}^n\!\hat\phi(t,\underline{x}), {}^n\!\hat\phi( t^\prime,
\underline{x}^\prime
 )\,]&=&0 \label{1Dphiphi}\\
 {[\,} {}^n\!\hat\pi( t,\underline{x} ), {}^n\!\hat\pi(
t^\prime,\underline{x}^\prime
 )\,]&=&0 \label{1Dpipi}\\
 {[\,} {}^n\!\hat\phi( t, \underline{x} ), {}^n\!\hat\pi( t^\prime,
\underline{x}^\prime
 )\,]&=&i\hbar \delta (t-t^\prime )\delta^3(
\underline{x} -\underline{x}^\prime ),
 \label{1Dphipi}
\end{eqnarray}
where, for each $t\in\mathR$, the fields ${}^n\!\hat\phi( t\,,
\underline{x})$ and ${}^n\!\hat\pi( t\,, \underline{x})$ are
defined on the space-like hypersurface characterised by the unit
length time-like vector $n$, and by the foliation parameter $t$.
In particular, the three-vector $\underline{x}$ in $
{}^n\!\hat\phi(t, \underline{x})$ or in ${}^n\!\hat\pi(
t,\underline{x})$, denotes a vector in this space. Note that the
pair $(t,\underline{x})$ can be used to identify a unique point
$X$ in spacetime, and hence to write $ {}^n\!\hat\phi(t,
\underline{x})$ as $ {}^n\!\hat\phi(X)$. The history algebra Eqs.
(\ref{1Dphiphi})--(\ref{1Dphipi}) can then be written in the more
covariant looking form
\begin{eqnarray}
{[}\,\hat\phi(X),\hat\phi(X')\,]&=&0  \label{QFTHAphiphi}\\
{[}\,\hat\pi(X),\hat\pi(X)\,]&=&0     \label{QFTHApipi}\\
{[}\,\hat\phi(X),\hat\pi(X')\,]&=&i\hbar\delta^4(X-X')
                                        \label{QFTHAphipi}
\end{eqnarray}
where we have dropped the $n$ superscript on the fields since the
algebra itself is $n$-independent. Of course, this does not stop
individual {\em representations\/} from depending on the
foliation vector $n$; indeed, as we shall see below, this is
precisely what happens.

In what follows we shall denote by $H_+:=\{n\in M\mid n\cdot n=1,
n^0>0\}$ the set of all unit length, forward pointing time-like
vectors on Minkowski spacetime $M$. We are using a metric
$\eta_{\mu\nu}$ on $M$ with the signature $(+,-,-,-)$; also we
use the notation $a\cdot b:=a^\mu b^\nu\eta_{\mu\nu}$ for any
four-vectors $a$ and $b$ in $M$.

It was shown in \cite{SavQFT1} that for each fixed $n$ in $H_+$
it is possible to find a representation of the history algebra
Eqs (\ref{1Dphiphi})--(\ref{1Dphipi}) on a Hilbert space ${\cal
H}_n$ with the property that the time-averaged energy exists as a
well-defined self-adjoint operator ${}^n\!\hat H$ (this is the
history analogue of an old theorem of Araki in the context of
canonical quantum field theory \cite{Araki60}). This operator
generates translations along the time-like direction $n$ and, as
such, is one of the generator of the {\em internal\/} Poincar\'e
group that exists for each $n$: full expressions for all these
generators are given in \cite{SavQFT1}.

One of the key questions for our present purposes is how the
external Poincar\'e group acts for each fixed choice of the
foliation vector $n$. The translation part should obviously act
in analogue to Eq.\ (\ref{ExtTTransl}) by taking $
{}^n\!\hat\phi(X)$ to $ {}^n\!\hat\phi(X+a)$ for any four-vector
$n$. Thus there should be an operator $U(a)$ such that
\begin{equation}
        U(a){}^n\!\hat\phi(X)U(a)^{-1}={}^n\!\hat\phi(X+a),
\end{equation}
with a similar action on ${}^n\!\hat\pi(X)$.

The  Lorentz subgroup of the Poincar\`e group is more interesting
since as well as acting on the spacetime points, it might also be
expected to act on the foliation vector $n$, and hence to take us
out of the Hilbert space ${\cal H}_n$. In \cite{SavQFT1} this
problem is solved by showing that even though the representations
of the field algebra (\ref{1Dphiphi})--(\ref{1Dphipi}) are
unitarily inequivalent for different choices of $n$, it is
nevertheless possible to construct the fields for all $n$ on a
{\em common\/} Fock space $\cal F$ (see Section
\ref{Sec:QHTScalarField} of the present paper for details). Hence
it is meaningful to look for a unitary operator $U(\Lambda)$ such
that, for all Lorentz transformations $\Lambda$,
\begin{equation}
    U(\Lambda){}^n\!\hat\phi(X) U(\Lambda)^{-1}=
                {}^{\Lambda n}\!\hat\phi(\Lambda X),
                \label{ULambdaaction}
\end{equation}
and similarly for ${}^n\!\hat\pi(X)$, where the operators are all
defined on $\cal F$. Of course, the operators $U(\Lambda)$ are
expected to form a unitary representation of the Lorentz group in
the sense that
\begin{equation}
    U(\Lambda')U(\Lambda)=U(\Lambda'\Lambda).    \label{Urepn}
\end{equation}

In the present paper we shall extend this formalism by quantising
the foliation vector $n$ itself. The main motivation for this
step is our belief that, when constructing the quantum history
theory of general relativity \cite{SavGR1}, it will be necessary
to include the spacetime foliation\footnote{In \cite{KK01} also,
the analogue of the foliation is a part of the postulated history
group, this time in the context of the Bosonic string.}  as a
genuine `history variable', and which must therefore be
represented by operators in the corresponding quantum theory. In
the context of our present discussion, the vector $n$ is the
Minkowskian analogue of a foliation of a general spacetime: hence
an investigation into what is meant by quantising $n$ is a
valuable precursor to the study of the quantisation of foliations
of a general spacetime.

As an introduction to the quantisation of $n$, it is useful to
return to the idea that, for each $n$, the theory is defined on a
Hilbert space ${\cal H}_n$, and to ask again how the external
Lorentz group might act. In these circumstances, Eq.\
(\ref{ULambdaaction}) is not meaningful since the operators
${}^n\!\hat\phi(X)$ and ${}^{\Lambda n}\!\hat\phi(\Lambda X)$ are
defined on different Hilbert spaces (${\cal H}_n$ and ${\cal
H}_{\Lambda n}$ respectively). The natural thing instead is to
seek a family of unitary intertwining operators
$U(n;\Lambda):{\cal H}_n\rightarrow{\cal H}_{\Lambda n}$ with the
property that
\begin{eqnarray}
U(n;\Lambda){}^n\!\hat\phi(X) ={}^{\Lambda n}\!\hat\phi(\Lambda X)
U(n;\Lambda)\\[2pt]
U(n;\Lambda){}^n\!\hat\pi(X) ={}^{\Lambda n}\!\hat\pi(\Lambda X)
U(n;\Lambda)
\end{eqnarray}
which can usefully be represented by the commutative diagram
\begin{equation}
\bundle{{\cal H}_n}{{}^n\!\hat\phi(X)}{{\cal H}_n}
\bundlemap{U(n;\Lambda)}{U(n;\Lambda)} %
\bundle{{\cal H}_{\Lambda n}}{{}^{\Lambda n}\!\hat\phi(\Lambda
X)}{{\cal H}_{\Lambda n}}
\end{equation}
and similarly for the operator ${}^n\!\hat\pi(X)$.

These operators $U(n;\Lambda):{\cal H}_n\rightarrow{\cal
H}_{\Lambda n}$ are expected to give a type of `representation'
of the external Lorentz group in the sense that, for all $n\in
H_+$ and for all Lorentz transformations $\Lambda$, we have
\begin{equation}
    U(\Lambda n; \Lambda')U(n;\Lambda)=U(n;\Lambda'\Lambda)
                        \label{UInt}
\end{equation}
which is the appropriate analogue of the genuine representation
Eq.\ (\ref{Urepn}). The specific form of Eq.\ (\ref{UInt})
follows by considering the commutative diagram
\begin{equation}
\bundle{{\cal H}_n}{{}^n\!\hat\phi(X)}{{\cal H}_n}
\bundlemap{U(n;\Lambda)}{U(n;\Lambda)} %
\bundle{{\cal H}_{\Lambda n}}{{}^{\Lambda n}\!\hat\phi(\Lambda
X)}{{\cal H}_{\Lambda n}}%
\bundlemap{U(\Lambda n,\Lambda')}{U(\Lambda n,\Lambda')}
\bundle{{\cal H}_{\Lambda'\Lambda n}}{{}^{\Lambda'\Lambda
n}\!\hat\phi(\Lambda'\Lambda X)}{{\cal H}_{\Lambda'\Lambda n}}
\end{equation}
whose outer square should equal the diagram
\begin{equation}
\bundle{{\cal H}_n}{{}^n\!\hat\phi(X)}{{\cal H}_n}
\bundlemap{U(n;\Lambda'\Lambda)}{U(n;\Lambda'\Lambda)} %
\bundle{{\cal H}_{\Lambda'\Lambda n}}{{}^{\Lambda'\Lambda
n}\!\hat\phi(\Lambda'\Lambda X)}{{\cal H}_{\Lambda'\Lambda n}}%
\end{equation}

We note that Eq.\ (\ref{UInt}) is the type of relation that
occurs naturally whenever we have a group $G$ that acts on some
$G$-set $X$, together with a family of operators $U_x(g)$, $x\in
X$, defined on vector spaces $V_x$, $x\in X$, and satisfying the
equation (cf.\ Eq.\ (\ref{UInt}))
\begin{equation}
        U(gx,g')U(x,g)=U(x,g'g)             \label{Ugx}
\end{equation}
for all $x\in X$ and $g,g'\in G$. There is also a version of Eq.\
(\ref{Ugx}) that uses a multiplier, but we shall not need that
here.

Mathematically speaking, the appropriate picture (for the
specific case of Eq.\ (\ref{UInt})) is a bundle of Hilbert spaces
${\cal H}_n$, $n\in H_+$, with base space $H_+$, in which the
action $n\mapsto \Lambda n$ of the external Lorentz group
$SO(3,1)$ on $H_+$ is lifted to the bundle by the maps
$U(n,\Lambda):{\cal H}_n\rightarrow{\cal H}_{\Lambda n}$; note
that Eq.\ (\ref{UInt}) is precisely the statement that the
operators $U(n,\Lambda)$ `cover' the action of $SO(3,1)$ on the
base space $H_+$.

Under these circumstances, it is natural to consider the new
Hilbert space formed by the {\em cross-sections\/} of this vector
bundle. However this Hilbert space is quite different from the
individual spaces ${\cal H}_n$, $n\in H_+$: in particular, the
foliation vector itself becomes an operator under the natural
action on a section $\Psi$ as
\begin{equation}
\{\hat n^\mu\Psi\}(n):=n^\mu\Psi(n)
\end{equation}
for all $n\in H_+$.

All this will be discussed properly in Section
\ref{Sec:QuFoliation}, but for the moment it suffices to
summarise our remarks above by saying that the mathematical
formalism for a history quantum field theory developed in
\cite{SavQFT1} itself suggests a natural way in which the
foliation vector could become a quantum operator.

Such a step is also understandable from a more conceptual
perspective. For we should recall that the consistent history
theory deals with `beables' (albeit contextualised by the choice
of a particular consistent set of histories) not observables.
Thus, in quantising $n$, we are not saying that the foliation is
something that is determined by nature---in particular, something
that must be observed---but rather that the existing history  QFT
formalism depends on the choice of $n$ in such an intrinsic way
that it is natural to formulate propositions about things in the
context\footnote{This also suggests that a topos approach could
be useful: we shall make a few remarks about this later on in the
paper.} of specifying $n$. And, as we have seen, one way of doing
this in a form that is coherent with respect to the action of the
external Lorentz group is to let $n$ become a quantum operator.

The challenge now arises of finding a proper theory of a
quantised foliation vector, and thereby justifying the rather
heuristic ideas presented above. In particular, in the spirit of
our approach to history theories, we must find the correct
extension of the history field algebra to include an operator
$\hat n_\mu$ and its conjugate variables.

We will address this issue in Section \ref {SubSec:multisymp} by
considering the {\em multi-symplectic\/} approach to a scalar
field theory. There is a relation between multi-symplectic
structures and history theory, and---as we shall see---in the
case of a scalar field, attempting to quantise the corresponding
multi-symplectic structure leads naturally to a quantised
foliation vector in the context of a history group whose Lie
algebra generators include $n$ and an appropriate set of conjugate
variables.

The plan of the paper is as follows. We start in Section
\ref{Sec:QHTScalarField} by summarising the results in
\cite{SavQFT1} for constructing the quantum history theory of a
free scalar field. Then in Section \ref{Sec:QuFoliation} we study
the main problem of quantising the foliation vector. We base the
first few steps in constructing the appropriate history algebra
on the discussion in Section \ref{SubSec:multisymp} of the
multi-symplectic formalism as applied to the relativistic scalar
field. The quantisation of this formalism is discussed in Section
\ref{Subsec:FirstSteps}, and this is completed in Section
\ref{CompletingHistoryAlgebra} where we apply group-theoretical
quantisation techniques to a classical system whose configuration
space is the set $H_+$ of all foliation vectors. Then in Section
\ref{Sec:Repns} we discuss the representations of this algebra,
and show how a particularly simple one reproduces the heuristic
ideas of a Hilbert bundle sketched above.

\section{The Quantum History Theory of a Scalar Field}
\label{Sec:QHTScalarField}
\subsection{The field operators}
The starting point for the construction in \cite{SavQFT1} of a
quantum history version of a free, scalar field is a Fock space
${\cal F}$ defined via annihilation and creation operators, $\hat
b(X)$ and $\hat b^\dagger(X)$ respectively, that satisfy the
commutation relations:
\begin{eqnarray}
&&{[\,}\hat b(X),\, \hat b( X^{\prime})\,] = 0                \\
&&{[\,}\hat b^\dagger(X),\, \hat b^\dagger(X^{\prime})\,] = 0 \\
&&{[\,}\hat b(X),\, \hat b^\dagger(X^{\prime})] =
            \hbar\delta^4(X -X^{\prime}).
\end{eqnarray}
This bosonic Fock space has (generalised) basis vectors
$\ket{X_1,X_2,\ldots, X_k}$ defined by
\begin{equation}
    \ket{X_1,X_2,\ldots, X_k}:=\hat b^\dagger(X_1)\hat b^\dagger(X_2)
    \ldots \hat b^\dagger(X_k)\ket{0}
\end{equation}
where $\ket{0}$ is the cyclic `vacuum' state of the Fock space.

On this Fock space $\cal F$, field operators for each $n\in H_+$
can be defined as
\begin{eqnarray}
 {}^n\!\hat\phi(X) &:=& {1\over\sqrt2} {{}^n\Gamma}^{-1/4} \Big(\hat b(X)+
  \hat b^{\dagger}(X)\Big)   \label{Def:nphib}\\
 {}^n\!\hat\pi(X) &:=& {1\over i\sqrt2} {{}^n\Gamma}^{1/4} \Big(\hat b(X)-
 \hat b^{\dagger}(X)\Big)  \label{Def:npib}
\end{eqnarray}
where ${}^n\Gamma$ is the elliptic, partial differential operator
on $L^2(\mathR^4)$ defined as
\begin{equation}
    {}^n\Gamma:=(\eta^{\mu\nu}-n^\mu
    n^\nu)\partial_\mu\partial_\nu+m^2
\end{equation}
where $m$ is the mass parameter in the theory. It is easy to
check that, for each foliation vector $n\in H_+$, the fields
${}^n\!\hat\phi(X)$ and ${}^n\!\hat\pi(X)$ defined in Eqs.\
(\ref{Def:nphib})--(\ref{Def:npib}) satisfy the history algebra
Eqs.\ (\ref{QFTHAphiphi})--(\ref{QFTHAphipi}). We note that, as
promised, these fields are all defined on the {\em same\/} space
$\cal F$ even though the associated representations of the
history algebra are unitarily inequivalent for different choices
of $n\in H_+$.

The time-averaged energy for each $n$ is represented by the
operator
\begin{eqnarray}
{}^n\!\hat H&=&{1\over 2}:\int
d^4X\{{}^n\!\hat\pi(X)^2+{}^n\!\hat\phi(X)\,{}^n\Gamma
                \,{}^n\!\hat\phi(X)\}:\label{Def:nH1}\\
&=&\int d^4X\,\hat b^\dagger(X)\sqrt{{}^n\Gamma}\, \hat b(X)
\end{eqnarray}
which is a well-defined self-adjoint operator. It is to guarantee
the existence of these operators for all $n\in H_+$ that the
basic fields ${}^n\!\hat\phi(X)$ and ${}^n\!\hat\pi(X)$ are
defined as they are in Eqs.\ (\ref{Def:nphib})--(\ref{Def:npib}).

\subsection{The external Poincar\'e group}
There is a natural unitary representation of the `external'
Poincar\'e group on the Fock space $\cal F$. This is defined in
the obvious way on the basic vectors $\ket{X_1,X_2,\ldots, X_k}$
as
\begin{eqnarray}
U(\Lambda)\ket{0}&:=&\ket{0}\\
U(\Lambda)\ket{X_1,X_2,\ldots, X_k}&:=&\ket{\Lambda X_1,\Lambda
                    X_2,\ldots, \Lambda X_k}\\[2pt]
U(a)\ket{0}&:=&\ket{0}\\
U(a)\ket{X_1,X_2,\ldots, X_k}&:=&\ket{X_1+a,X_2+a,\ldots, X_k+a}.
\end{eqnarray}
This induces the action on the annihilation operators of
\begin{eqnarray}
U(\Lambda)\hat b(X)U(\Lambda)^{-1}&=&\hat b(\Lambda X)\\
U(a)\hat b(X) U(a)^{-1} &=& \hat b(X+a),
\end{eqnarray}
and similarly for the creation operators $\hat b^\dagger(X)$. It
is straightforward to show that, as anticipated, the basic field
operators ${}^n\!\hat\phi(X)$ and ${}^n\!\hat\pi(X)$ transform as
\begin{eqnarray}
U(\Lambda){}^n\!\hat\phi(X)U(\Lambda)^{-1}&=&
        {}^{\Lambda n}\!\hat\phi(\Lambda X)\label{Transfnphi}\\
U(\Lambda){}^n\!\hat\pi(X)U(\Lambda)^{-1}&=&
        {}^{\Lambda n}\!\hat\pi(\Lambda X)
        \label{Transfnpi}\\[2pt]
U(a){}^n\!\hat\phi(X)U(a)^{-1}&=& {}^n\!\hat\phi(X+a)\\
U(a){}^n\!\hat\pi(X)U(a)^{-1}&=& {}^n\!\hat\pi(X+a).
\end{eqnarray}

We note that it is possible to define  another set of fields by
\begin{eqnarray}
\hat\Phi(X)&:=&{1\over\sqrt2}(\hat b(X)+\hat b^\dagger(X))\\
\hat\Pi(X)&:=&{1\over i\sqrt2}(\hat b(X)-\hat b^\dagger(X))
\end{eqnarray}
which satisfy the basic history field algebra Eqs.\
(\ref{QFTHAphiphi})--(\ref{QFTHAphipi}). Under the action of the
external Poincar\'e group, we have
\begin{eqnarray}
U(\Lambda)\hat \Phi(X)U(\Lambda)^{-1}&=&\hat \Phi(\Lambda X)\\
U(a)\hat \Phi(X) U(a)^{-1} &=& \hat \Phi(X+a),
\end{eqnarray}
and similarly for the conjugate variable $\hat\Pi(X)$.

The role of these `covariant' fields in the theory is intriguing.
The relation of $\hat\Phi(X)$  to the fields ${}^n\!\hat\phi(X)$
suggests strongly that the former should be thought of as the
history analogue of the {\em Newton-Wigner\/} field (which, in
standard quantum field theory, creates and annihilates localised
{\em particle\/} states). However we note that---in the history
theory---$\hat\Phi(X)$ is a genuine scalar field, whereas in
standard quantum field theory the Newton-Wigner field transforms
in a non-covariant way.

The formal explanation of this difference lies in the way the
internal and external times interface with each other in the
history theory. In particular, the history field $\hat\Phi(X)$ is
a `Schr\"odinger  picture' object in the sense that it does not
carry any dynamical information. On the other hand, the remarks
above about the standard Newton-Wigner field apply in the
Heisenberg picture: in the history case, this would involve
invoking the second, internal time.

\section{Quantising the Foliation Vector}
\label{Sec:QuFoliation}
\subsection{The multi-symplectic formalism for a scalar field}
\label{SubSec:multisymp} One might be tempted to construct the
classical history theory for a scalar field by positing the
Poisson bracket algebra (cf.\ Eqs.\
(\ref{QFTHAphiphi})--(\ref{QFTHAphipi}))
\begin{eqnarray}
{\{}\,\phi(X),\phi(X')\,\}&=&0  \label{ClassHAphiphi}\\
{\{}\,\pi(X),\pi(X')\,\}&=&0     \label{ClassHApipi}\\
{\{}\,\phi(X),\pi(X')\,\}&=&\delta^4(X-X')
                                        \label{ClassHAphipi}
\end{eqnarray}
which has the advantage of appearing to be manifestly covariant
under the action of the external Poincar\'e group (on the
assumption that $\phi$ and $\pi$ are scalar fields). However,
this covariance is deceptive in the sense that the conjugate
variable $\pi(X)$ has no clear {\em physical\/} meaning; not
least because the actual field momentum for a physical system is
manifestly foliation dependent: {\em i.e.\/} it means the
momentum along some specified time-like direction $n$.

This problem is circumscribed in the approach summarised in
Section \ref{Sec:QHTScalarField} since the {\em
representations\/} of the quantum algebra Eqs.\
(\ref{QFTHAphiphi})--(\ref{QFTHAphipi}) are manifestly
$n$-dependent, and indeed an explicit $n$-label becomes attached
to both the scalar field and its conjugate momentum via Eqs.\
(\ref{Def:nphib})--(\ref{Def:npib}). However, these explicit forms
are chosen so that the {\em quantum\/} average-energy operator
${}^n\!\hat H$ exists, and to some extent therefore this leaves
open the question of the structure of the underlying {\em
classical\/} history theory. We shall now address this issue with
the aid of some ideas drawn from an, apparently, quite different
scheme: namely, the multi-symplectic formalism.

The multi-symplectic formalism arose from attempts to modify the
standard classical {\em canonical\/} formalism so that it would
be manifestly covariant under the appropriate group of spacetime
transformations \cite{Marsdenatal}. In the case of a scalar field
on Minkowski spacetime $M$, the idea is to introduce a pair of
space-time fields $\phi(X)$ and $\pi_\mu(X)$ where, physically,
for any vector $V$, $V^\mu\pi_\mu(X)$ can be interpreted as the
field momentum along the space-time direction $V^\mu$. Then, for
each choice of foliation vector $n$, there is defined the Poisson
bracket
\begin{equation}
\{F,G\}_n(\phi,\pi):=\int_M d^4X\left({\delta
F\over\delta\phi(X)} {\delta G\over\pi_\mu(X)}- {\delta
G\over\delta\phi(X)} {\delta F\over\pi_\mu(X)}\right)n_\mu
\label{Def:PBFGn}
\end{equation}
where $F$ and $G$ are functionals of $\phi$ and $\pi$. By this
means, a family of symplectic structures is introduced, and the
whole system is manifestly covariant under an action of the
Poincar\'e group in which (i) $\phi$ and $\pi_\mu$ transform as
genuine space-time objects in the obvious way; and (ii) the
symplectic structure labeled by a foliation vector $n$ is
transformed into that labeled by $\Lambda n$ for all Lorentz
transformations $\Lambda$.

The nature of this covariance is particularly clear if we look at
the basic Poisson brackets that follow from Eq.\
(\ref{Def:PBFGn}):
\begin{eqnarray}
\{\phi(X),\,\phi(X')\}_n&=&0                      \label{PBnphiphi}\\
\{\pi_\mu(X),\,\pi_\nu(X')\}_n&=&0                \label{PBnpipi}\\
\{\phi(X),\,\pi_\mu(X')\}_n&=&n_\mu\delta^4(X-X') \label{PBnphipi}
\end{eqnarray}
where $X$ and $X'$ are points in Minkowski spacetime $M$.

As remarked above, the multi-symplectic formalism was developed
in the context of standard canonical theory. However---in so far
as they are space-time objects---we could clearly think of $\phi$
and $\pi_\mu$ as classical {\em history\/} fields, and try to
develop a history theory based on Eqs.\
(\ref{PBnphiphi})--(\ref{PBnphipi}) instead of Eqs.\
(\ref{ClassHAphiphi})--(\ref{ClassHAphipi}). As a mathematical
possibility, this makes good sense. However, we should emphasise
that, physically speaking, the history interpretation of the
multi-symplectic formalism is quite different from the standard
one.

For example, a frequent comment in the literature on the
multi-symplectic formalism is that the basic Poisson brackets
Eqs.\ (\ref{PBnphiphi})--(\ref{PBnphipi}) are not compatible with
the equations of motion. But viewed as a history theory, this is
no longer the case since the equations of motion are now to be
associated with the introduction of the `internal' time label.
This is closely related to the fact that, from a history
perspective, the fields $\phi$ and $\pi_\mu$ are the classical
analogue of {\em Schr\"odinger\/} picture objects, and are used
in a temporal logic sense as the carriers of propositions about
the history of the system; they are {\em not\/} dynamical fields.

\subsection{First steps to the quantum history algebra}
\label{Subsec:FirstSteps} We must now address the question of the
quantum analogue of the parametrised (by $n$) family of Poisson
brackets given in Eqs.\ (\ref{PBnphiphi})--(\ref{PBnphipi}). It
is noteworthy that very little has been said in the literature on
the multi-symplectic formalism about quantising such Poisson
bracket relations, and by hindsight we can understand why: it is
only in the context of a quantum {\em history\/} theory---for
example, the consistent history theory---that the quantisation
makes any physical sense.

If we approach quantisation in the traditional way of replacing
Poisson brackets with operator commutators, then the first issue
is how to handle the $n$-subscript that appears on the left hand
side of the equations (\ref{PBnphiphi})--(\ref{PBnphipi}).
Attaching a subscript to an operator commutator does have any
{\em a priori\/} meaning other than, perhaps, to indicate
different representations of an algebra, and one is tempted
therefore to postulate the simple algebra (from now on we set
$\hbar=1$)
\begin{eqnarray}
{[}\,\hat \phi(X),\hat\phi(X')\,]&=&0           \\
{[}\,\hat \pi_\mu(X),\hat\pi_\nu(X')\,]&=&0     \\
{[}\,\hat\phi(X),\hat\pi_\mu(X')\,]&=&in_\mu\delta^4(X-X')
\label{cnumbern}
\end{eqnarray}
with the understanding that the physically appropriate
representation may depend on $n$.

However, the quantity $n_\mu$ now appears as a fixed $c$-number,
and the manifest Poincar\'e covariance is lost. For example, one
would like to postulate an action of the external Lorentz group
of the form
\begin{eqnarray}
U(\Lambda)\hat\phi(X)U(\Lambda)^{-1}&=&\hat\phi(\Lambda X)
                                \label{LTphi}\\
U(\Lambda)\hat\pi_\mu(X)U(\Lambda)^{-1}&=&\Lambda_\mu{}^\nu
        \hat\pi_\nu(\Lambda X)  \label{LTpi}
\end{eqnarray}
but this is incompatible with the right hand side of Eq.\
(\ref{cnumbern}) because, since $n_\mu$ is a $c$-number, we have
$U(\Lambda)n_\mu U(\Lambda)^{-1}=n_\mu$. The obvious resolution
of this problem is to make $n_\mu$ itself into an {\em
operator\/}, with the algebra
\begin{eqnarray}
{[}\,\hat \phi(X),\hat\phi(X')\,]&=&0       \label{nhatphiphi}\\
{[}\,\hat \pi_\mu(X),\hat\pi_\nu(X')\,]&=&0 \label{nhatpipi}    \\
{[}\,\hat\phi(X),\hat\pi_\mu(X')\,]&=&i \hat n_\mu\delta^4(X-X')
\label{nhatphipi}
\end{eqnarray}
and to augment the transformations Eqs.\
(\ref{LTphi})--(\ref{LTpi}) with
\begin{equation}
U(\Lambda)\hat n_\mu U(\Lambda)^{-1}=\Lambda_\mu{}^\nu\hat n_\nu
                                    \label{LTn}
\end{equation}
so that the whole set is now compatible.

We now have four main tasks:
\begin{enumerate}
\item Extend Eqs.\ (\ref{nhatphiphi})--(\ref{nhatphipi}) to a
complete history theory; in particular we must discuss the form
of the conjugate variables to the quantised foliation vector
$\hat n_\mu$.

\item Find a physically appropriate representation of the
extended algebra.

\item Show how dynamics is implemented in this scheme. In
particular, how the idea arises of a second, `internal' time and
associated internal Poincar\'e group.

\item Give a physical interpretation of the algebra.
\end{enumerate}

Of course, these different issues are closely related. For
example, the average-energy operator for the system could be
anticipated to be
\begin{equation}
\hat H={1\over 2}:\int d^4X\,\left\{(\hat n^\mu\hat\pi_\mu(X))^2
+(\hat n^\mu\hat n^\nu-\eta^{\mu\nu})\partial_\mu\hat\phi(X)
\partial_\nu\hat\phi(X)+m^2\hat\phi(X)^2\right\}: \label{Def:Hnhat}
\end{equation}
which should be compared with the expression in Eq.\
(\ref{Def:nH1}) for a fixed $n$-vector. Note that there is no
longer an $n$-superscript on $\hat H$: there is now just a single
operator. It is natural, therefore, to seek to fix the
representation of the final history algebra by requiring that the
expression in Eq.\ (\ref{Def:Hnhat}) exists as a genuine
(essentially) self-adjoint operator.

\subsection{Completing the history algebra}
\label{CompletingHistoryAlgebra} \paragraph{The conjugate
variables to $n$.} The next step is to consider the conjugate
variables for the foliation vector. The key observation in this
context is that, before quantisation, the vector $n$ is of unit
length in the sense that
\begin{equation}
    n\cdot n:=n^\mu n^\nu\eta_{\mu\nu}=1,    \label{ndotn=1}
\end{equation}
and time-like. It seems appropriate that the quantisation of $n$
should preserve these constraints, but this requirement is
incompatible with the obvious commutator algebra
\begin{equation}
    [\,\hat p^\mu,\hat n_\nu\,]=-i\delta_\nu^\mu
\end{equation}
since the conjugate $\hat p^\mu$ would then generate translations
in $\hat n_\mu$, and these do not preserve the constraints.

What we are faced with is the quantisation of a system whose
classical configuration space is not a vector space but rather
the hyperboloid $H_+:=\{n\in\mathR^4\mid n^\mu n_\mu=1, n^0>0\}$
in $\mathR^4$, which can be viewed as a non-compact version of the
three-sphere $S^3$. The quantisation of systems whose
configuration spaces are not vector spaces was discussed at
length in \cite{LesHouches84} which, in particular, contains a
detailed description of the quantisation of a system whose
classical configuration space is an $n$-sphere. The conclusion
was that the appropriate canonical group is not the standard Weyl
group (that is associated with the normal canonical commutation
relations) but rather the euclidean group
$SO(n+1)\copyright\mathR^{n+1}$ where $\copyright$ denotes the
semi-direct product.

The same general discussion applies in the present case with the
hyperboloid $H_+$ as configuration space. The result is that the
appropriate history group for the foliation variable is the
semi-direct product $SO(3,1)\copyright\mathR^4$, with the Lie
algebra relations
\begin{eqnarray}
    {[}\, \hat n_\alpha,\,\hat n_\beta\,]&=&0
                                            \label{HAnn}\\
    {[}\,\hat p^{\alpha\beta},\,\hat p^{\gamma\delta}\,]
    &=&i(\eta^{\alpha\gamma}\hat p^{\beta\delta}-
    \eta^{\beta\gamma}\hat p^{\alpha\delta} +
    \eta^{\beta\delta}\hat p^{\alpha\gamma} -
    \eta^{\alpha\delta}\hat p^{\beta\gamma})\label{HApp}\\
    {[}\,\hat n_\alpha,\, \hat p^{\beta\gamma}\,]&=&
      i(\delta_\alpha^\beta\hat n^\gamma -
        \delta_\alpha^\gamma\hat n^\beta)   \label{HAnp}
\end{eqnarray}
where $\hat p^{\alpha\beta}=-\hat p^{\beta\alpha}$.

We note that:
\begin{enumerate}
\item[i)] Eq.\ (\ref{HAnn}) shows that the variables $\hat n_\alpha$
span the Lie algebra of the abelian group $\mathR^4$.

\item[ii)] Eq.\ (\ref{HApp}) shows that the conjugate variables $\hat
p^{\alpha\beta}$ satisfy the Lie algebra  of the Lorentz group
$SO(3,1)$.

\item[iii)] Eq.\ (\ref{HAnp}) reflects the semi-direct structure given by
the action of $SO(3,1)$ on $\mathR^4$.
\end{enumerate}
This group-theoretic scheme works because $\eta^{\mu\nu}\hat
n_\mu\hat n_\nu$ is a Casimir operator for the algebra in Eqs.\
(\ref{HAnn})--(\ref{HAnp}). Hence it is meaningful to look for a
representation in which $\eta^{\mu\nu}\hat n_\mu\hat n_\nu$ has
the constant value 1, thus maintaining compatibility with the
classical constraint in Eq.\ (\ref{ndotn=1}).

Of course $SO(3,1)\copyright\mathR^4$ is nothing but the familiar
Poincar\'e group. But this should not be confused with either the
internal or the external Poincar\'e groups to which we have
referred earlier: the present group has arisen as a direct result
of quantising the foliation vector $n_\mu$.

\paragraph{Completing the history algebra.} We must now try to complete the
history algebra by considering the cross commutators between the
pair $\hat\phi(X)$, $\hat\pi_\mu(X)$, and the pair $\hat n_\mu$,
$\hat p^{\alpha\beta}$. As a first step we take the commutator of
both sides of Eq.\ (\ref{nhatphipi}) with $\hat p^{\alpha\beta}$,
then use the Jacobi identity on the left hand side, and the
commutator Eq.\ (\ref{HAnp}) on the right hand side, to give
\begin{equation}
[\,\hat\phi(X),[\,\hat p^{\alpha\beta},\hat\pi_\mu(X')]] +
[\,\hat\pi_\mu(X'),[\,\hat\phi(X),\hat p^{\alpha\beta}]]=
(\delta_\mu^\alpha\hat n^\beta-\delta_\mu^\beta\hat n^\alpha)
\delta^4(X-X'). \label{JI1}
\end{equation}

It is natural to think of $\hat\phi(X)$ and $\hat n_\mu$ as
disjoint configuration variables, which suggests that
\begin{equation}
    [\,\hat\phi(X),\hat n_\mu\,]=0=[\,\hat\pi_\alpha(X),\hat
    n_\mu\,],     \label{HAphin}
\end{equation}
and for this reason it is arguably also natural to assume that
$[\hat\phi(X),\,\hat p^{\alpha\beta}\,]=0$. We note that a more
general possibility is
\begin{equation}
[\,\hat\phi(X),\hat p^{\alpha\beta}\,]= ia(\hat
n^\alpha\hat\pi^\beta(X) -\hat n^\beta\hat\pi^\alpha(X))
                \label{HAphip}
\end{equation}
for some real constant $a$. However, this is rather ugly in the
sense that the right hand side of Eq.\ (\ref{HAphip}) is a
non-linear function of our basic fields, and from now on we shall
assume that $a=0$.

We note that, by virtue of Eq.\ (\ref{nhatpipi}) and the
assumption in Eq.\ (\ref{HAphin}), even if the commutator in Eq.\
(\ref{HAphip}) is non-zero, it does not contribute to the left
hand side of Eq.\ (\ref{JI1}). Thus, even if $a\neq 0$, the
obvious choice for the commutator $[\,\hat
p^{\alpha\beta},\hat\pi_\mu(X)\,]$ is
\begin{equation}
    [\,\hat p^{\alpha\beta},\hat\pi_\mu(X)]=-i(\delta_\mu^\alpha
    \hat\pi^\beta(X)-\delta_\mu^\beta\hat\pi^\alpha(X)).
\end{equation}

In summary, the entire history algebra is postulated to be as
follows:
\begin{eqnarray}
{[}\,\hat \phi(X),\hat\phi(X')\,]&=&0       \label{FHAphiphi}\\
{[}\,\hat \pi_\mu(X),\hat\pi_\nu(X')\,]&=&0 \label{FHApipi}    \\
{[}\,\hat\phi(X),\hat\pi_\mu(X')\,]&=&i
    \hat n_\mu\delta^4(X-X')                \label{FHAphipi}\\
{[}\, \hat n_\alpha,\,\hat n_\beta\,]&=&0   \label{FHAnn}\\
{[}\,\hat p^{\alpha\beta},\,\hat p^{\gamma\delta}\,]
    &=&i(\eta^{\alpha\gamma}\hat p^{\beta\delta}-
    \eta^{\beta\gamma}\hat p^{\alpha\delta} +
    \eta^{\beta\delta}\hat p^{\alpha\gamma} -
    \eta^{\alpha\delta}\hat p^{\beta\gamma})\label{FHApp}\\
{[}\,\hat n_\alpha,\, \hat p^{\beta\gamma}\,]&=&
      i(\delta_\alpha^\beta\hat n^\gamma -
        \delta_\alpha^\gamma\hat n^\beta)   \label{FHAnp}\\
{[}\,\hat\phi(X),\hat n_\alpha\,]&=&0        \label{FHAphin}\\
{[}\,\hat\pi_\mu(X),\hat n_\alpha\,]&=&0     \label{FHApin}\\
{[}\,\hat\phi(X),\hat p^{\alpha\beta}\,]&=& 0 \label{FHAphip}\\
{[}\,\hat\pi_\mu(X),\,\hat p^{\alpha\beta}\,]
    &=&i(\delta_\mu^\alpha \hat\pi^\beta(X)-
        \delta_\mu^\beta\hat\pi^\alpha(X)).  \label{FHApip}
\end{eqnarray}
It is easy to check that the Jacobi identities are satisfied for
this algebra.

\paragraph{An ansatz for the operator $\hat\pi_\mu(X)$.} At this point we
note that Eqs.\ (\ref{FHAphipi}) and (\ref{FHAphin}) imply that
\begin{equation}
{[}\,\hat\phi(X),\hat\pi_\mu(X')-\hat n_\mu\, \hat
n_\nu\hat\pi^\nu(X')\,]=0 \label{phiTranspi}
\end{equation}
in a representation in which $\hat n\cdot \hat n=1$. For an
arbitrary value of this Casimir operator we have instead
\begin{equation}
{[}\,\hat\phi(X),\,\hat n\cdot \hat n\,\hat\pi_\mu(X')-\hat
n_\mu\, \hat n_\nu\hat\pi^\nu(X')\,]=0.
\end{equation}

Equation (\ref{phiTranspi}) suggests that $\hat\pi_\mu(X)-\hat
n_\mu\, \hat n_\nu\hat\pi^\nu(X)$ might be a function of
$\hat\phi(X)$: indeed, this statement is necessarily true if the
algebra generated by the spacetime fields $\hat\phi(X)$ is
assumed to be a maximal commutative subalgebra of the history
algebra. One natural possibility is to set
\begin{equation}
\hat\pi_\mu(X)-\hat n_\mu\, \hat n_\nu\hat\pi^\nu(X)=0,
\end{equation}
which suggests that $\hat\pi_\mu(X)$ can be defined using a
single `master' field $\hat\pi(X)$ as
\begin{equation}
    \hat\pi_\mu(X):=\hat n_\mu\hat\pi(X). \label{Def:pimu}
\end{equation}
We shall discuss this option at some length below. Note that it
is compatible with the supposed commutator $[\,\hat
\pi_\mu(X),\hat\pi_\nu(X')\,]=0$ if we postulate that
$[\,\hat\pi(X),\hat\pi(X')\,]=0$. It is also compatible with the
remaining commutators in Eqs.\ (\ref{FHAphiphi})--(\ref{FHApip})
that involve $\hat\pi_\mu(X)$.

A natural generalisation of the definition Eq.\ (\ref{Def:pimu})
of the operator $\hat\pi_\mu(X)$ in terms of a single
$\hat\pi(X)$ is
\begin{equation}
\hat\pi_\mu(X):=\hat n_\mu\hat\pi(X)+b(\partial_\mu\hat\phi(X) -
                \hat n_\mu \hat n\cdot\partial\hat\phi(X))
                    \label{pimu=longtransf}
\end{equation}
for some real constant $b$. Bearing in mind that (assuming that
$\hat n\cdot\hat n=1$)
\begin{equation}
    \hat n^\mu(\partial_\mu\hat\phi(X) -
                \hat n_\mu \hat n\cdot\partial\hat\phi(X))\equiv 0
\end{equation}
we see that Eq.\ (\ref{pimu=longtransf}) can be viewed as the
decomposition of $\hat\pi_\mu(X)$ into a `longitudinal' part
$\hat n_\mu\hat\pi(X)$ and a `transverse part'
$\partial_\mu\hat\phi(X) -\hat n_\mu \hat
n\cdot\partial\hat\phi(X)$, with the implication in particular
that the transverse part is essentially the spatial derivatives
of the field $\hat\phi(X)$. There are several attractive features
to assuming Eq.\ (\ref{pimu=longtransf}). However, it does have
the implication that
\begin{equation}
    [\,\hat\pi_\mu(X),\hat\pi_\nu(X')\,]=2ib(\hat
    n_{(\nu}\partial_{\mu)}-\hat n_\mu\hat n_\nu \hat n\cdot\partial)
        \delta^4(X-X')              \label{newppipi}
\end{equation}
where the partial derivatives on the right hand side are with
respect to the $X$ label. This would mean making a change in the
postulated commutator in Eq.\ (\ref{FHApipi}).

\subsection{The external and internal Poincar\'e groups}
\paragraph{The action of the external Poincar\'e group.}
There is a natural automorphism of the complete history algebra
Eqs.\ (\ref{FHAphiphi})--(\ref{FHApip}) by the external
Poincar\'e group, which we might hope could be unitarily
implemented  as an extension of Eqs.\ (\ref{LTphi})--(\ref{LTpi})
and Eq.\ (\ref{LTn}):
\begin{eqnarray}
U(\Lambda)\hat\phi(X)U(\Lambda)^{-1}&=&\hat\phi(\Lambda X)
                                \label{ELTphi}\\
U(\Lambda)\hat\pi_\mu(X)U(\Lambda)^{-1}&=&\Lambda_\mu{}^\nu
        \hat\pi_\nu(\Lambda X)  \label{ELTpi}\\
U(\Lambda)\hat n_\mu U(\Lambda)^{-1}&=&
            \Lambda_\mu{}^\nu\hat n_\nu  \label{ELTn}\\
U(\Lambda)\hat p^{\alpha\beta} U(\Lambda)^{-1}&=&
\Lambda^\alpha{}_\mu \Lambda^\beta{}_\nu\hat
p^{\mu\nu}\label{ELTp}
\end{eqnarray}
and with the translations acting as
\begin{eqnarray}
U(a)\hat\phi(X)U(a)^{-1}&=&\hat\phi(X+a)\label{ETrphi}\\
U(a)\hat\pi_\mu(X)U(a)^{-1}&=&\hat\pi_\mu(X+a)\label{ETrpi}\\
U(a)\hat n_\mu U(a)^{-1}&=&\hat n_\mu\label{ETrn}\\
U(a)\hat p^{\alpha\beta}U(a)^{-1}&=&\hat p^{\alpha\beta}.
                                    \label{ETrp}
\end{eqnarray}

\paragraph{The internal Poincar\'e group.} The situation with the internal
Poincar\'e group is interesting. As remarked above, we expect the
average-energy operator of the system to be
\begin{equation}
\hat H={1\over 2}:\int d^4X\,\left\{(\hat n^\mu\hat\pi_\mu(X))^2
+(\hat n^\mu\hat n^\nu-\eta^{\mu\nu})\partial_\mu\hat\phi(X)
\partial_\nu\hat\phi(X)+m^2\hat\phi(X)^2\right\}:
                    \label{Def:Hnhat2}
\end{equation}
However, in this situation, $\hat H$ is not defined with respect
to any {\em particular\/} foliation (unlike, for example, the
time-averaged energy in Eq.\ (\ref{Def:nH1})), and hence it
cannot be identified as the time-like component of a four-vector
$\hat P_\mu$ in any obvious way.

The resolution of this issue is as follows. In the original form
of history quantum field theory, summarised in Section \ref
{Sec:QHTScalarField}, there is a four-vector operator ${}^n\!\hat
P_\mu$ that is related to the associated time-averaged energy
operator ${}^n\!\hat H$ by
\begin{equation}
    {}^n\!\hat P_\mu :=n_\mu {}^n\!\hat H+
        \int d^4X\left(\partial_\mu{}^n\!\hat\phi(X)-n_\mu
        n\cdot\partial\;{}^n\!\hat\phi(X)\right){}^n\!\hat\pi(X),
                            \label{Def:nPmu}
\end{equation}
and where we note that
\begin{equation}
    n^\mu\int d^4X\left(\partial_\mu{}^n\!\hat\phi(X)-n_\mu
n\cdot\partial\;{}^n\!\hat\phi(X)\right)
 {}^n\!\hat\pi(X)\equiv 0. \label{nPmutrans=0}
\end{equation}
Thus the `$n$-longitudinal' part of ${}^n\!\hat P_\mu$ is
${}^n\!\hat H\equiv n^\mu\, {}^n\!\hat P_\mu$, whereas the
`$n$-transverse' part is $\int
d^4X\left(\partial_\mu\hat\phi(X)-n_\mu
n\cdot\partial\,\hat\phi(X)\right){}^n\!\hat\pi(X)$.

In the present case, where $n$ is quantised, the expression in
Eq.\ (\ref{Def:nPmu}) suggests that we define the translation
generators of the internal Poincar\'e group by
\begin{equation}
    {}^{\rm int}\!\hat P_\mu:=\hat n_\mu \hat H+
        \int d^4X\left(\partial_\mu\hat\phi(X)-\hat n_\mu
        \hat n\cdot\partial\,\hat\phi(X)\right){}^n\!\hat\pi(X)
\end{equation}
where $\hat H$ is defined in Eq.\ (\ref{Def:Hnhat2}). The
remaining generators of the internal Poincar\'e group can be
defined in a similar way using the expressions given in
\cite{SavQFT1} where the foliation vector $n$ is fixed.

We note that, according to Eqs.\ (\ref{ELTphi})--(\ref{ELTpi}),
under the action of the external Lorentz group, the generators of
the translations of the internal Poincar\'e group transform as
\begin{equation}
U(\Lambda)\,{}^{\rm int}\!\hat P_\mu\, U(\Lambda)^{-1}=
        \Lambda_\mu{}^\nu \,{}^{\rm int}\!\hat P_\nu
\end{equation}
whereas, for a fixed $n$ we have
\begin{equation}
U(\Lambda)\,{}^n\!\hat P_\mu\, U(\Lambda)^{-1}=
        \Lambda_\mu{}^\nu \,{}^{\Lambda n}\!\hat P_\nu.
\end{equation}

\paragraph{The internal time.}
In the context of the discussion above of the internal Poincar\'e
group, it is clear that one way in which a second, internal time
variable $s$ could enter the formalism is by the definition of a
`Heisenberg picture' field $\hat\phi(X;s)$ as
\begin{equation}
    \hat\phi(X;s):=e^{is\hat H}\hat\phi(X)e^{-is\hat H}.
                                \label{Def:phi(X;s)}
\end{equation}
We see that, in this approach, there is now a {\em single\/}
extra time variable $s$---for each choice of a foliation vector
$n$---and this is not associated with any particular foliation
vector. However, it is still true that the interpretation of the
formalism must be such that $s$ automatically has the correct
meaning in the correct context.

However, this is not the only option. For example, it is arguably
more natural to have a separate internal time variable $s(n)$ for
each value of $n\in H_+$, and such that $s(n)\geq 0$ for all $n$.
In the quantum case, an operator $s(\hat n)$ can be defined using
the spectral theorem for the self-adjoint operator $\hat n$, and
then we can define (c.f. Eq.\ (\ref{Def:Hnhat2}))
\begin{equation}
\hat H[s]:={1\over 2}:\int d^4X\,s(\hat n)\left\{(\hat
n^\mu\hat\pi_\mu(X))^2 +(\hat n^\mu\hat
n^\nu-\eta^{\mu\nu})\partial_\mu\hat\phi(X)
\partial_\nu\hat\phi(X)+m^2\hat\phi(X)^2\right\}:
                    \label{Def:H[s]}
\end{equation}
This suggests defining an associated `generalised Heisenberg
picture' object $\hat\phi(X;s]$ (c.f. Eq.\ (\ref{Def:phi(X;s)}))
as
\begin{equation}
    \hat\phi(X;s]:=e^{i\hat H[s]}\hat\phi(X)e^{-i\hat H[s]}
                                \label{Def:phi(X;s]}
\end{equation}
where the brackets in $\hat\phi(X;s]$ serve to remind us that
$\hat\phi$ is a function of the spacetime point $X$, but a {\em
functional\/} of the function $s:H_+\rightarrow
\{0\}\cup\mathR_+$.

\section{Representations of the History Algebra}
\label{Sec:Repns}
\subsection{The Hilbert bundle construction}
From what has been said above, it is clear that one way of
satisfying the history algebra Eqs.\
(\ref{FHAphiphi})--(\ref{FHApip}) would be to have a single
`master' momentum field $\hat\varpi(X)$, and then to define
\begin{equation}
    \hat\pi_\mu(X):=\hat n_\mu\hat\varpi(X)
\end{equation}
with the assumption that $\hat n_\mu(X)$ commutes with
$\hat\varpi(X)$ so that there are no operator-ordering problems.
This gives us the simpler algebra
\begin{eqnarray}
{[}\,\hat\phi(X),\,\hat\phi(X')\,]&=&0      \label{SHAPhiPhi}\\
{[}\,\hat\varpi(X),\,\hat\varpi(X')\,]&=&0      \label{SHAPiPi}\\
{[}\,\hat\phi(X),\,\hat\varpi(X')\,]&=&i\delta^4(X-X')\label{SHAPhiPi}\\
{[}\, \hat n_\alpha,\,\hat n_\beta\,]&=&0   \label{SHAnn}\\
{[}\,\hat p^{\alpha\beta},\,\hat p^{\gamma\delta}\,]
    &=&i(\eta^{\alpha\gamma}\hat p^{\beta\delta}-
    \eta^{\beta\gamma}\hat p^{\alpha\delta} +
    \eta^{\beta\delta}\hat p^{\alpha\gamma} -
    \eta^{\alpha\delta}\hat p^{\beta\gamma})\label{SHApp}\\
{[}\,\hat n_\alpha,\, \hat p^{\beta\gamma}\,]&=&
      i(\delta_\alpha^\beta\hat n^\gamma -
        \delta_\alpha^\gamma\hat n^\beta)   \label{SHAnp}
\end{eqnarray}
with all other commutators vanishing. Of course, this is just the
direct sum of the field algebra Eqs.\
(\ref{SHAPhiPhi})--(\ref{SHAPhiPi}) with the algebra Eqs.\
(\ref{SHAnn})--(\ref{SHAnp}). Note that the commutator
${[}\,\hat\pi_\mu(X),\,\hat p^{\alpha\beta}\,]
=i(\delta_\mu^\alpha \hat\pi^\beta(X)-
\delta_\mu^\beta\hat\pi^\alpha(X))$ in Eq.\ (\ref{FHApip}) need
no longer be assumed since it is implied now by the commutation
relation ${[}\,\hat n_\alpha,\, \hat p^{\beta\gamma}\,] =
i(\delta_\alpha^\beta\hat n^\gamma - \delta_\alpha^\gamma\hat
n^\beta)$ in Eq.\ (\ref{SHAnp}).

We anticipate that the key average-energy operator (that will
eventually be associated with translations along the internal
time direction) is (cf.\ Eq.\ (\ref{Def:Hnhat2}))
\begin{equation}
    \hat H:={1\over 2}:\int d^4X\left\{\hat\varpi(X)^2+(\hat n^\mu\hat
    n^\nu-\eta^{\mu\nu})\partial_\mu\hat\phi(X)\partial_\nu\hat\phi(X)+m^2
    \hat\phi(X)^2\right\}:   \label{Def:masterH}
\end{equation}
in which case the main task is to find a representation of the
history algebra Eqs.\ (\ref{SHAPhiPhi})--(\ref{SHAnp}) in which
Eq.\ (\ref{Def:masterH}) exists as a genuine self-adjoint
operator.

We proceed as follows. For each $n\in H_+$ we construct the
Hilbert space ${\cal H}_n$ that carries operators
${}^n\hat\phi(X)$, ${}^n\hat\pi(X)$ that satisfy the history
algebra Eqs.\ (\ref{QFTHAphiphi})--(\ref{QFTHAphipi}):
\begin{eqnarray}
{[}\,{}^n\!\hat\phi(X),{}^n\!\hat\phi(X')\,]&=&0  \label{nQFTHAphiphi}\\
{[}\,{}^n\!\hat\pi(X),{}^n\!\hat\pi(X)\,]&=&0     \label{nQFTHApipi}\\
{[}\,{}^n\!\hat\phi(X),{}^n\!\hat\pi(X')\,]&=&i\delta^4(X-X')
                                       \label{nQFTHAphipi}
\end{eqnarray}
and with the property that the average-energy operator in Eq.
(\ref{Def:nH1})
\begin{equation}
    {}^n\!\hat H:={1\over 2}:\int d^4X\left\{\hat{}^n\!\hat \pi(X)^2
    +(n^\mu n^\nu -\eta^{\mu\nu})\partial_\mu{}^n\!\hat\phi(X)\,
    \partial_\nu{}^n\!\hat\phi(X)+m^2\,\,
    {}^n\!\hat\phi(X)^2\right\}:   \label{Def:nH2}
\end{equation}
exists as a genuine self-adjoint operator. The $n$-superscripts
on the fields in Eqs.\ (\ref{nQFTHAphiphi})--(\ref{nQFTHAphipi})
serve to indicate that we have chosen the representation of the
abstract history algebra Eqs.\
(\ref{QFTHAphiphi})--(\ref{QFTHAphipi}) in which this operator
${}^n\!\hat H$ exists. In fact, as we know from the work in
\cite{SavQFT1}, for all $n\in H_+$ these fields can be
constructed on the same abstract Fock space even though the
corresponding representations of the history field algebra are
unitarily inequivalent for different $n$. However, for our
present purposes, it is clearer if we continue to refer to the
Hilbert space on which ${}^n\!\hat H$ exists as ${\cal H}_n$.

We now link up with the heuristic ideas in the Introduction by
constructing a Hilbert bundle whose base space is the hyperboloid
$H_+$, and in which the fiber over each $n\in H_+$ is defined to
be the Hilbert space ${\cal H}_n$.\footnote{We note that the
$SO(3,1)$ subgroup of the history group ({\em i.e.}, the part
associated with the $\hat n_\mu$ variables) acts transitively on
$H_+$ with stability group $SO(3)$, so that $H_+\simeq
SO(3,1)/SO(3)$. Thus we have the principle bundle
\begin{equation}
    SO(3)\longrightarrow SO(3,1)\longrightarrow
    SO(3,1)/SO(3)\simeq H_+:=\{n\in\mathR^4\mid n\cdot n=1,
    n^0>0\}.                         \label{SO31bundle}
\end{equation}
This suggests that we could try using a non-trivial
representation of $SO(3)$ to `twist' the fibers of the Hilbert
bundle. However, we shall not explore that option here.} The
Hilbert space of our history theory is then defined to be the
direct integral
\begin{equation}
    {\cal H}:=\int^\oplus_{H_+}{\cal H}_n\,d\mu(n).
\end{equation}
Here $d\mu(n)$ is the usual $SO(3,1)$-invariant measure on the
hyperboloid $H_+$: it is just the standard measure used in normal
quantum field theory, but now applied to $n$-space rather than
momentum space.

The vectors in this direct-integral Hilbert space $\cal H$ are
defined to be the cross-sections of the Hilbert bundle: {\em
i.e.}, maps $\Psi:H_+\rightarrow \bigcup_{n\in H_+} {\cal H}_n$
with the property that $\Psi(n)\in{\cal H}_n$ for all $n\in H_+$.
The inner product between a pair of such cross-sections $\Psi_1$
and $\Psi_2$ is defined as
\begin{equation}
    \langle\Psi_1,\Psi_2\rangle:=
    \int_{H_+}d\mu(n)\langle\Psi_1(n),\Psi_2(n)\rangle_{{\cal
    H}_n}
\end{equation}
where $\langle\,\,,\,\rangle_{{\cal H}_n}$ denotes the inner
product in the Hilbert space fiber ${\cal H}_n$.

Of course, if we make the specific identification of each Hilbert
space ${\cal H}_n$ with the Fock space $\cal F$, as summarised in
Section \ref{Sec:QHTScalarField}, then the new Hilbert space
$\cal H$ can be viewed as  the vector space of all measurable
functions $\Psi:H_+\rightarrow{\cal F}$ with the inner product
\begin{equation}
\langle\Psi_1,\Psi_2\rangle:=
    \int_{H_+}d\mu(n)\langle\Psi_1(n),\Psi_2(n)\rangle_{\cal F}.
\end{equation}
Note that there is a natural cyclic `ground' state which is
defined to be the cross-section $\Omega$ such that, for all $n\in
H_+$,
\begin{equation}
    \Omega(n):=\ket{0}_n
\end{equation}
where $\ket{0}_n$ is the ground state of the average-energy
operator ${}^n\!\hat H$ in the Hilbert-space ${\cal H}_n$.

\subsection{The field and foliation operators}
\paragraph{The field operators.}
The next step is to define the history field operators
$\hat\phi(X)$ and $\hat\varpi(X)$ on ${\cal
H}:=\int^\oplus_{H_+}{\cal H}_n\,d\mu(n)$ as follows:
\begin{eqnarray}
    \{\hat\phi(X)\Psi\}(n)&:=&{}^n\!\hat\phi(X)\Psi(n)\label{Def:Phi}\\
    \{ \hat\varpi(X)\Psi\}(n)&:=&{}^n\!\hat\pi(X)\Psi(n)    \label{Def:Pi}
\end{eqnarray}
for all $n\in H_+$. These equations are meaningful since the
vectors $\Psi(n)$, $n\in H_+$, in the right hand sides belong to
the Hilbert space ${\cal H}_n$ on which the field operators
${}^n\!\hat\phi(X)$ and ${}^n\!\hat\pi(X)$ are defined. In other
words, the maps $n\mapsto {}^n\!\hat\phi(X)$ and $n\mapsto
{}^n\!\hat\pi(X)$ define fields of operators over the base space
$H_+$, and are hence well-defined\footnote{Of course, to do this
rigorously one needs to discuss the domains of the various
operators concerned, but we shall not dwell on such niceties
here.} operators on the direct integral $\int^\oplus_{H_+}{\cal
H}_n\, d\mu(n)$.

It is clear that the operators defined by Eqs.\ (\ref{Def:Phi})
and (\ref{Def:Pi}) satisfy the history algebra Eqs.\
(\ref{SHAPhiPhi})--(\ref{SHAPhiPi}). For example,
\begin{equation}
\{\hat\phi(X)\hat\varpi(X')\Psi\}(n)={}^n\!\hat\phi(X)\{\hat\varpi(X')\Psi\}
(n)
            ={}^n\!\hat\phi(X)\,{}^n\!\hat\pi(X')\Psi(n)
\end{equation}
and similarly
\begin{equation}
\{\hat\varpi(X')\hat\phi(X)\Psi\}(n)={}^n\!\hat\pi(X')\{\hat\phi(X)\Psi\}(n)
            ={}^n\!\hat\pi(X')\,{}^n\!\hat\phi(X)\Psi(n)
\end{equation}
so that, for all $n\in H_+$,
\begin{eqnarray}
\left\{[\,\hat\phi(X),\hat\varpi(X')\,]\Psi\right\}(n)&=&
 [\,{}^n\!\hat\phi(X),{}^n\!\hat\pi(X')\,]\Psi(n)\nonumber\\
                    &=&i\delta^4(X-X')\Psi(n)
\end{eqnarray}
which means that (modulo subtleties about domains) we have the
basic history field commutator
$[\,\hat\phi(X),\hat\varpi(X')\,]=i\delta^4(X-X')$.

Note that, if we exploit the fact that the Hilbert spaces can all
be identified with the same Fock space $\cal F$, then using the
definitions in Eqs.\ (\ref{Def:nphib})--(\ref{Def:npib}), we can
further write
\begin{eqnarray}
    \{\hat\phi(X)\Psi\}(n)&:=&{}^n\!\hat\phi(X)\Psi(n)
    ={1\over\sqrt2}{{}^n\Gamma}^{-1/4} \Big(\hat b(X)+
  \hat b^{\dagger}(X)\Big)\Psi(n)\label{Def:Phi2}\\
    \{\hat\varpi(X)\Psi\}(n)&:=&{}^n\!\hat\pi(X)\Psi(n)
    ={1\over i\sqrt2} {{}^n\Gamma}^{1/4} \Big(\hat b(X)-
 \hat b^{\dagger}(X)\Big)\Psi(n).    \label{Def:Pi2}
\end{eqnarray}
Here the operator $\hat b(X)$ (and similarly for $\hat
b^\dagger(X)$) is defined as the constant field of operators over
$H_+$ obtained by identifying each fiber of the Hilbert bundle
with the Fock space $\cal F$.

\paragraph{The foliation operators.}
The operators that represent the foliation vector are easy to
define in the Hilbert space $\int^\oplus_{H_+}{\cal
H}_n\,d\mu(n)$. Specifically:
\begin{equation}
    \{\hat n_\mu\Psi\}(n):=n_\mu\Psi(n) \label{Def:n}
\end{equation}
and
\begin{equation}
    \{\hat p^{\alpha\beta}\Psi\}(n):=
    i\left\{n_\alpha{\partial\over\partial n_\beta}-
            n_\beta{\partial\over\partial
            n_\alpha}\right\}\Psi(n). \label{Defp}
\end{equation}
Note that, strictly speaking, if the history states are
considered as sections of the Hilbert bundle, then the right hand
side of Eq.\ (\ref{Defp}) involves taking the difference between
vectors belonging to different Hilbert-space fibers, and hence it
is only meaningful if there is a {\em connection\/} in the
bundle. However, this is not a problem in our case since the
fibers ${\cal H}_n$, $n\in H_+$, can all be identified with the
basic Fock space $\cal F$, and this is assumed to have been done
when writing Eq.\ (\ref{Defp}).

\paragraph{The time-averaged energy operator.}
The natural way of defining a time-averaged energy operator is to
exploit the fact that, on each Hilbert space fiber ${\cal H}_n$,
$n\in H_+$, the operator ${}^n\!\hat H$ defined in Eq.\
(\ref{Def:nH2}) exists, and represents the time-averaged value of
the energy for that particular foliation. Thus we can
define\footnote{As usual, to be fully rigorous we should worry
about domains of essential self-adjointness for these operators.}
an operator $\hat H$ by
\begin{equation}
    \{\hat H\Psi\}(n):={}^n\!\hat H\Psi(n)
\end{equation}
for all $n\in H_+$. Note that, as anticipated in Eq.\
(\ref{Def:masterH}), the operator thus defined can be written in
terms of the basic history fields as
\begin{equation}
    \hat H:={1\over 2}:\int d^4X\left\{\hat\varpi(X)^2+(\hat n^\mu\hat
    n^\nu-\eta^{\mu\nu})\partial_\mu\hat\phi(X)\partial_\nu\hat\phi(X)+m^2
    \hat\phi(X)^2\right\}:   \label{masterH}
\end{equation}
The remaining generators of the internal Poicar\'e group can be
defined in an analogous way.

\paragraph{The internal time function.}
If an internal time function is introduced as in Eq.\
(\ref{Def:H[s]}), then the action of the operator $\hat H[s]$ on
a section $\Psi$ of the Hilbert bundle is
\begin{equation}
        \{\hat H[s]\Psi\}(n):=s(n)\,{}^n\!\hat H\Psi(n),
\end{equation}
which shows clearly the sense in which $s(n)$ is the internal
time associated with the foliation vector $n$.

This suggests an interesting application of the ideas discussed
in \cite{BI00} of possible uses of topos theory in quantum
gravity and quantum theory. In particular, one might try to view
the construction above as being, rather than of a bundle, instead
of a a sheaf of Hilbert spaces over the base space $H_+$, which
is now construed as the category of `contexts' in which
assertions about the history system are to be made.

By viewing our construction as an object in the topos of sheaves
over $H_+$, we can exploit the existence in any topos of both
external and internal views: `external' in the sense of how
things look from the perspective of normal mathematics; and
`internal' in the sense of how things look from the perspective
of the mathematical structure based on the topos itself. In
particular, when viewed externally, the time function $n\mapsto
s(n)$ appears precisely as that: {\em i.e.}, a function. On the
other hand, when viewed internally it corresponds to a real
number in the topos of sheaves over $H_+$. Thus what we have
called the `internal time {\em function\/}' is just a {\em real
number\/} when viewed internally in the topos. We intend to
devote a future paper to the general question of the ways in
which topos ideas can be productively applied to history theory.

\subsection{The external Poincar\'e group}
The key step in constructing a representation of the external
Lorentz group in the Hilbert space ${\cal H}$ of cross-sections
is to have a family of intertwining operators $U(n;\Lambda):
{\cal H}_n\rightarrow {\cal H}_{\Lambda n}$ that satisfy the
conditions given in Eq.\ (\ref{UInt}):
\begin{equation}
    U(\Lambda n; \Lambda')U(n;\Lambda)=U(n;\Lambda'\Lambda).
                        \label{UInt2}
\end{equation}
Indeed, the conditions in Eq.\ (\ref{UInt2}) mean precisely that
the intertwining operators $U(n;\Lambda)$ `cover' ({\em i.e.,}
act coherently with respect to) the action of $SO(3,1)$ on the
base space $H_+$ of the Hilbert bundle.

In these circumstances, for each $\Lambda\in SO(3,1)$, we can
define an operator $W(\Lambda):{\cal H}\rightarrow {\cal H}$ by
\begin{equation}
    \{W(\Lambda)\Psi\}(n):=
    U(\Lambda^{-1}n;\Lambda)\Psi(\Lambda^{-1}n)
    \label{Def:WLambda}
\end{equation}
for all $n\in H_+$. This is clearly unitary since
\begin{eqnarray}
    \langle W(\Lambda)\Psi,\,W(\Lambda)\Psi\rangle_{\cal H}&=&
    \int_{H_+}d\mu(n)\, \langle
\{W(\Lambda)\Psi\}(n),\,\{W(\Lambda)\Psi\}(n)
                \rangle_{{\cal H}_n}\nonumber\\
                &=&\int_{H_+}d\mu(n)\, \langle
                U(\Lambda^{-1}n;\Lambda)\Psi(\Lambda^{-1}n),
                \,U(\Lambda^{-1}n;\Lambda)\Psi(\Lambda^{-1}n)\rangle_{{\cal
                H}_n}\nonumber\\
 &=& \int_{H_+}d\mu(n)\, \langle \Psi(\Lambda^{-1}n),\,\Psi(\Lambda^{-1}n)
    \rangle_{{\cal H}_{\Lambda^{-1}n}}\nonumber\\
 &=& \int_{H_+}d\mu(n)\, \langle \Psi(n),\,\Psi(n)\rangle_{{\cal
 H}_n}\nonumber\\
 &=& \langle\Psi,\Psi\rangle_{\cal H}
\end{eqnarray}
where have used (i) the assumed unitarity of the intertwining
operators $U(n;\Lambda): {\cal H}_n\rightarrow {\cal H}_{\Lambda
n}$; and (ii)  the invariance of the measure $d\mu$ on $H_+$
under the action of $SO(3,1)$.

To see that $W(\Lambda)$ defined in Eq.\ (\ref{Def:WLambda})
satisfies the group law we compute as follows:
\begin{eqnarray}
\{W(\Lambda_2)W(\Lambda_1)\Psi\}(n)&=&
U(\Lambda_2^{-1}n;\Lambda_2)
\left(\{W(\Lambda_1)\Psi\}(\Lambda_2^{-1}n)\right)  \nonumber\\
&=&U(\Lambda_2^{-1}n;\Lambda_2)U(\Lambda_1^{-1}\Lambda_2^{-1}n;\Lambda_1)
    \Psi(\Lambda_1^{-1}\Lambda_2^{-1}n). \label{Wgrouplaw}
\end{eqnarray}
But, from Eq.\ (\ref{UInt2}) we have
\begin{equation}
U(\Lambda_2^{-1}n;\Lambda_2)U(\Lambda_1^{-1}\Lambda_2^{-1}n;\Lambda_1)
= U(\Lambda_1^{-1}\Lambda_2^{-1}n;\Lambda_2\Lambda_1)=
U((\Lambda_2\Lambda_1)^{-1}n;\Lambda_2\Lambda_1)
\end{equation}
and hence, for all $n\in H_+$,
\begin{eqnarray}
\{W(\Lambda_2)W(\Lambda_1)\Psi\}(n)&=&
U((\Lambda_2\Lambda_1)^{-1}n;\Lambda_2\Lambda_1)
\Psi((\Lambda_2\Lambda_1)^{-1}n)\nonumber\\
&=&\{W(\Lambda_2\Lambda_2)\Psi\}(n)
\end{eqnarray}
as is required to give a representation of the Lorentz group.

As was mentioned earlier, in our particular case, the existence
of intertwining operators $U(n;\Lambda): {\cal H}_n\rightarrow
{\cal H}_{\Lambda n}$ satisfying Eq.\ (\ref{UInt2}) is
demonstrated rather easily by exploiting the fact that the Hilbert
spaces ${\cal H}_n$, $n\in H_+$, can all be identified naturally
with the same Fock space generated by creation and annihilation
operators $\hat b(X)^\dagger$ and $\hat b(X)$. Indeed, as
discussed earlier, we simply get operators $U(\Lambda):{\cal
F}\rightarrow {\cal F}$ which in themselves give a representation
of the external Lorentz group, and which satisfy Eqs.\
(\ref{Transfnphi})--(\ref{Transfnpi}).

The translation subgroup of the external Poincar\'e group is
easier to define since the translations do not act on $H_+$. Thus
we have the simple definition
\begin{equation}
    \{W(a)\Psi\}(n):={}^n\!U(a)\Psi(n)\makebox{\ \ $\forall n\in H_+$},
\label{Def:Wa}
\end{equation}
where ${}^n\!U(a)$ denotes the operators of the translation
subgroup of the external Poincar\'e group in ${\cal H}_n$.

\section{Conclusions}
We have shown how the discussion in \cite{SavQFT1} of a history
version of scalar quantum field theory can be augmented in such a
way as to include the quantisation of the unit-length, time-like
vector $n$ that determines the Lorentzian foliation of Minkowski
spacetime. The Hilbert bundle construction that we employed was
motivated by: (i) a heuristic discussion of the role of the
external Lorentz group in the existing history quantum field
theory \cite{SavQFT1}; and (ii) a more technical discussion of a
specific representation of the extended history algebra obtained
from the multi-symplectic representation of classical scalar
field theory. In the latter context it should be remarked that
there exist representations of this algebra other than the simple
one given here---the significance of such representations is a
subject for future research.

The construction of a Hilbert bundle over $H_+:=\{n\in M\mid
n\cdot n=1, n^0>0\}$ is a natural idea at a technical level, but
it is also interesting from a conceptual perspective. For
example, the direct integral representation of the history Hilbert
space---together with the postulated non-dependence of the
average energy operator on the variables conjugate to $\hat
n^\mu$---suggests that we have a type of history analogue of
what, in ordinary quantum theory, would be regarded as a system
with continuous super-selection sectors labelled by $n\in H_+$.
But in a `neo-realist' theory such as consistent histories, the
role of super-selection sectors is somewhat different from that
which arises in an instrumentalist theory such as standard quantum
mechanics.

However, the main motivation behind the present paper is to
present certain mathematical techniques that can be proved useful
when quantising the spacetime foliations that are expected to
arise in a history version of general relativity. This important
issue in the history approach to quantum gravity, is something to
which we shall return in a later paper.

\bigskip\noindent
{\large {\bf Acknowledgements}}

\noindent Support by the EPSRC in form of grant GR/R36572 is
gratefully acknowledged.

\end{document}